\documentclass[apj]{emulateapj}

\usepackage{multirow}
\usepackage{ulem}
\newcommand{\gf}{ }
\def\rhessi{{\textit{RHESSI}}}
\def\mw{{microwave}}

\def\gs{{gyrosynchrotron}}

\def\peak{{_{\rm peak}}}
\def\plume{{_{\rm plume}}}

\shorttitle{Flare's Large-scale Plume}
\shortauthors{Fleishman et al.}

\begin{document}


\title{A Large-scale Plume in an X-Class Solar Flare }

\author{Gregory D. Fleishman$^1$}

\author{Gelu M. Nita$^1$}

\author{Dale E. Gary$^1$}

\altaffiliation{$^1$Physics Department, Center for Solar-Terrestrial Research, New Jersey Institute of Technology
Newark, NJ, 07102-1982}

\begin{abstract}
Ever-increasing multi-frequency imaging of solar observations suggests that solar flares often involve more than one magnetic fluxtube. Some of the fluxtubes are closed, while others can contain open field. The relative proportion of nonthermal electrons among those distinct loops is highly important for understanding the energy release, particle acceleration, and transport. The access of nonthermal electrons to the open field is further important as the open field facilitates the solar energetic particle (SEP) escape from the flaring site, and thus controls the SEP fluxes in the solar system, both directly and as seed particles for further acceleration. The large-scale fluxtubes are often filled with a tenuous plasma, which is difficult to detect in either EUV or X-ray wavelengths; however, they can dominate at low radio frequencies, where a modest component of nonthermal electrons can render the source optically thick and, thus, bright enough to be observed. Here we report detection of a large-scale `plume' at the impulsive phase of an X-class solar flare, SOL2001-08-25T16:23, using multi-frequency radio data from Owens Valley Solar Array. To quantify the flare spatial structure, we employ 3D modeling utilizing force-free-field extrapolations from the line-of-sight SOHO/MDI magnetograms with our modeling tool GX$\_$Simulator. We found that a significant fraction of the nonthermal electrons accelerated at the flare site low in the corona escapes to the plume, which contains both closed and open field. We propose that the proportion between the closed and open field at the plume is what determines the SEP population escaping into interplanetary space.
\end{abstract}


\section{Introduction}


Obtaining a complete picture of a solar flare requires detailed knowledge of particle acceleration, the partition of these particles among various magnetic structures (loops, jets, etc) involved in the flaring, as well as their transport, precipitation, and escape. Currently, the most detailed quantitative information is available from the thick-target hard X-ray (HXR) emission from the down-precipitating population of nonthermal electrons accelerated in flares. This thick-target HXR emission is routinely observed from the chromospheric footpoints of flares by the \textit{Reuven Ramaty High Energy Solar Spectroscopic Imager} \citep[\rhessi,][]{lin2002}. The spectrum of the footpoint HXR emission can be interpreted in terms of precipitating electron flux, which, under certain assumptions about the nonthermal electron transport in the flaring loop, can then be associated with the nonthermal electron population in the acceleration region. The so-inferred properties of the accelerated component depend heavily on the validity of the adopted mode of the nonthermal electron transport.

In some cases, detections of acceleration regions have been reported. In the HXR domain, the acceleration region is inferred from the  presence of above-the-loop-top sources \citep{masuda_etal_1994} and to coronal sources in some partly occulted flares \citep{krucker_lin_2008, krucker_etal_2010}, although this has been put into question by \citet{2014AAS...22410404H}. In addition, acceleration region detections using microwave data have been reported \citep{Fl_etal_2011, Fl_etal_2013}, which were made possible by careful joint analysis of both X-ray and \mw\ data, often augmented by 3D modeling \citep{Fl_etal_2016, Fl_Xu_etal_2016}. The \mw\ gyrosynchrotron emission is known to be often dominated by a nonthermal electron component trapped in a coronal magnetic loop \citep{Melnikov_1994,Bastian_etal_1998, Meln_Magun_1998, 2000ApJ...531.1109L, Kundu_etal_2001, melnikov_etal_2002}, although there are cases where this trapped component is weak or nonexistent. In such cases the direct contribution from the acceleration region can dominate \citep{Fl_etal_2016_narrow}.

Once accelerated in the impulsive phase of a flare, nonthermal electrons will leave the acceleration region and naturally collect in any magnetic reservoir to which they have access. The partition of nonthermal electrons among different regions involved in a flare is difficult to study with any single data set, primarily because most of the available imaging instruments in both X-ray and \mw\ domains have a very limited dynamic range; thus, one or more weaker sources may not be detected in the presence of a strong one. This is why there are very few studies of the nonthermal electron partitions among distinct flaring loops and they necessarily include 3D modeling and multiple data sets, which, to a certain extent, compensate for the limited dynamic ranges of individual instruments \citep[e.g.,][and Glesener \& Fleishman 2017]{Fl_etal_2016}.

Of particular importance are large-scale magnetic structures, which likely make a ``bridge'' between a relatively compact site of the flare energy release/particle acceleration and remote sites and/or interplanetary medium, thus offering ``escape routes'' for solar energetic particles (SEPs). Although these escape routes can sometimes be associated with jets visible in EUV and/or soft X-rays (SXRs), the jet plasma must be dense enough to be detected in either EUV or SXR ranges; if tenuous, such escape routes can be undetectable for EUV and SXR instruments. Here the microwave data can help greatly, given that even a relatively small number of nonthermal electrons can render a magnetized volume filled with these particles optically thick at relatively low frequencies and, thus, bright enough to dominate the spectrum. Indeed, the gyrosynchrotron opacity $\tau\approx L\varkappa$, where $L$ is the source depth along the line of sight (LOS) and $\varkappa$ is the absorption coefficient at a given frequency, rises towards lower frequencies $f$ as a power-law function with a large index, $\varkappa\propto f^{-\alpha}$, $\alpha\sim3-8$ \citep[see, e.g.,][for an order-of-magnitude estimate]{Dulk_1985}. In the optically thick case the brightness temperature from a given LOS is defined by ``effective'' energy $\langle E \rangle$ of the electrons responsible for emission at the given frequency, $T_{B}=T_{eff}\equiv \langle E \rangle / k_B$, where $k_B$ is the Boltzmann constant. In the Rayleigh-Jeans regime, always valid in the radio domain, the flux density of the optically thick emission is fully specified by three variables: emission frequency $f$, effective (or brightness) temperature $T_{eff}$, and the source area $A$: $F\propto (k_B f^2/c^2) \int T_{B} d \Omega \propto f^2 T_{eff} A$. This immediately tells us that having a large flux density of low-frequency radio emission implies a correspondingly large source area. For example, having a similar flux density at 10~GHz and 1~GHz, which is often observed in large flares \citep{1994SoPh..152..409L} and even in modest flares \citep{Fl_etal_2016}, requires that the area of the latter one is by a factor of 100 larger than that of the former one, other conditions being equal. In practice, this factor can be even greater, given that the effective temperature typically rises with frequency. This implies that the low-frequency \mw\ gyrosynchrotron emission can effectively be used to probe large-scale structures associated with flares in the solar corona, which hitherto have been largely unrecognized, but which may play an essential role in SEP production.

\begin{figure}
    \centering
    \includegraphics[width=0.98\columnwidth]{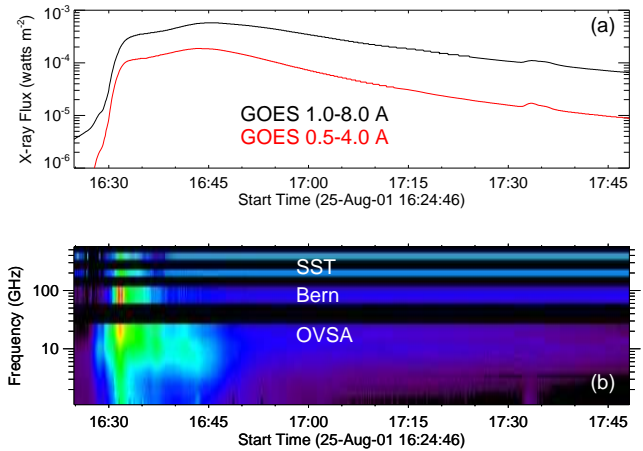}\\
    \caption{\label{f_20010825_overview} Overview of the SOL2001-08-25T16:23 solar flare. (a) GOES soft X-ray light curves. (b) Radio emission dynamic spectrum at 1--405 GHz combined from OVSA data at 1--18~GHz, Bern data at 89.4~GHz, and SST data at 212 \& 405~GHz.  }
\end{figure}

In this paper we consider a strong X5.3 flare that occurred on 25 Aug 2001, which has an exceptionally broad spectral coverage in the radio domain, 1--405~GHz. We analyse the radio emission from the impulsive phase of this flare with an emphasis on the low-frequency emission, employing both  imaging from Owens Valley Solar Array \citep[OVSA,][]{Gary_Hurford_1994} and  3D modeling. We find that the low-frequency radio emission comes from a large-scale ``plume'' associated with the flare, which may serve as an SEP escape route that does not show up in other wavelengths, because the physical parameters that allow the plume to be bright at the low radio frequencies also leave it undetectable at other wavelengths.

%
%
%
%

\section{Observations }

The X5.3 class solar flare that occurred at $\sim$16:23~UT on 2001-Aug-25, in AR~9591 (S17$^\circ$E34$^\circ$), was one of the strongest flares during solar cycle 23. Indeed, a large white light flare was observed \citep{2003ApJ...595..483M}, which also produced a significant flux of neutrons at Earth \citep{2003ICRC....6.3179W}, along with strong sub-THz emission \citep{2004SoPh..223..181R}. Various aspects of this flare have been discussed in numerous papers
\citep[e.g.,][]{2006SoSyR..40..153L, 2006SoSyR..40..104K, 2013PASJ...65S...4G}.

Here we specifically concentrate on the microwave imaging data obtained with OVSA, along with spectral radio data obtained with OVSA and other instruments \citep{2004SoPh..223..181R}.
The spectral data, Figure~\ref{f_20010825_overview}, consist of the total power (TP) data from OVSA at 40 frequencies roughly logarithmically spaced from 1--18~GHz, complemented by TP Bern data at 89.4~GHz and Solar Submillimeter-wave Telescope (SST) data at 212 and 405~GHz \citep[see][for details of the high-frequency data]{2004SoPh..223..181R}. This exceptionally broad spectral coverage, 1--405~GHz, is extremely important for this powerful event, given that the spectral peak frequency varied from a few GHz up to a few dozens GHz over the course of the flare. In particular, the peak frequency was well outside the OVSA spectral range at the main flare phase, so the higher-frequency measurements were needed to constrain the spectral shape of the \mw\ emission.

The \mw\ flux density was exceptionally strong: it reached values around $10^5$~sfu at 89.4~GHz, $\sim4\times 10^4$~sfu at 18~GHz, and almost 7500~sfu at 1.2~GHz. Such a high flux density of the \gs\ emission at such a low frequency is a direct indication that the source area (at the low frequencies) must be exceptionally large. Indeed, for the optically thick emission we have \citep{Fl_etal_2016}

\begin{equation}
\label{Eq_Thick_Flux}
  F_I=F_{\rm LCP}+ F_{\rm RCP}\simeq $$$$ 12~[{\rm sfu}] \left(\frac{f}{1~{\rm GHz}}\right)^{2}\left(\frac{T_{eff}}{10^7~{\rm K}}\right)\left(\frac{A}{10^{20}~{\rm cm}^2}\right);
\end{equation}
thus, having $F_I \sim 7500$~sfu at $\sim1$~GHz requires $A\sim 2\times10^{20}$~cm$^2$ even if we assume a very high effective temperature $T_{eff} \sim 3\times10^9$~K. That  source area implies a linear size around $3'$, which is comparable with the typical size of an entire active region.

To directly check this expectation, we need imaging information on the \mw\ sources.
The OVSA imaging capability and the adopted calibration scheme described elsewhere \citep{2013ApJ...769L..11L, Fl_etal_2015, Fl_etal_2016_narrow} have the ability to image solar flares at many adjacent frequencies between $\sim$~1--15~GHz, which is a significant advantage compared with any other solar-dedicated radio instrument available during solar cycle 23. Unfortunately, there are also limitations and disadvantages of OVSA  for imaging, such as the most severe of which are dissimilar sizes of the antennas and their correspondingly different fields of view, relatively small number of the antenna baselines available and, thus, undersampling of the $uv$ points needed for high dynamic-range imaging, and a necessarily complicated calibration scheme, which sometimes fails to produce reliable data for imaging.

\begin{figure}\centering
\includegraphics[width=0.98\columnwidth]{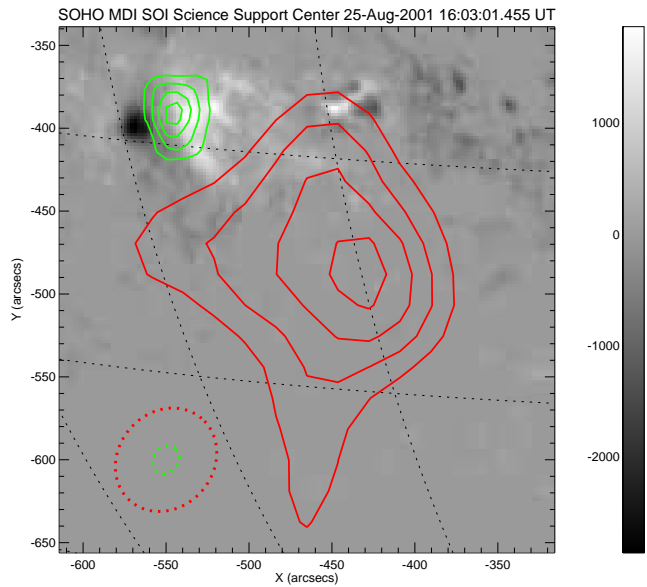}\\
\vspace{0.5in}
\caption{\label{f_mdi_OVSA} Preflare line-of-sight SOHO/MDI magnetogram (16:03:01~UT) and OVSA snapshot ($\sim$16:32:31~UT) images at 6.2~GHz (green) and 1.2--2~GHz frequency synthesis (red). Dotted ovals show the OVSA beams at the corresponding frequencies. The animated version of the figure (available from the on-line ApJ paper, when published) shows the evolution of the radio sources over the impulsive peak of the burst. Note very fast apparent motion of the low-frequency source compared with the high-frequency source, which stays at roughly same location all the time.    
 }
\end{figure}

Fortunately, the OVSA data are reasonably good for imaging in the case of the 25-Aug-2001 flare; the \mw\ sources are resolved at most of the considered frequencies (the observed source size is notably larger than the beam size). We attempted the imaging at various frequencies and frequency ranges and over various times, of which a representative subset of results is illustrated in Figure~\ref{f_mdi_OVSA}, summarized as follows:
\begin{enumerate}
  \item the source area increases towards lower frequencies, especially below $\sim$3~GHz;
  \item the \mw\ sources at high frequencies, $\gtrsim5$~GHz, project onto the core of the AR, where the magnetic field is strong;
  \item the \mw\ sources at low frequencies, $\lesssim3$~GHz, are strongly displaced compared with the core of the AR (and with the high-frequency \mw\ sources).
  \item this displacement holds for the entire flare duration, but the low-frequency source location notably evolves during the flare; see the animated version of Figure~\ref{f_mdi_OVSA}.
\end{enumerate}

Observations \# 1 \& 2 are expected from a qualitative consideration of the radio spectrum: we already noted that the source area must increase strongly toward lower frequencies, while having a high-frequency spectral peak does require a strong magnetic field at the source and, thus, the high-frequency source must project onto the AR core. However, observations \# 3 \& 4 look striking given that neither the low-frequency source displacement relative to the AR core nor its spatial evolution  are required by other data. On the other hand, it is not excluded because no strong magnetic field is needed to form the optically thick \gs\ emission at the low frequencies, $f\sim 1-3$~GHz. In fact, a relatively weak magnetic field of about 1--10~G \citep[cf.][]{1994SoPh..152..409L, Fl_etal_2016} is favored by the presumably high brightness temperature \citep{Dulk_1985} at the low frequencies. To reconcile if the displacement, along with other observations, is consistent with the context magnetic field data, we resort to 3D modeling using our modeling tool GX Simulator \citep{Nita_etal_2015}, similar to that performed by \citet{Kuznetsov_Kontar_2015, Fl_etal_2013, Fl_Xu_etal_2016, Fl_etal_2016}.

\section{3D Modeling of Radio Emission}

The 3D modeling of a flare \citep{Nita_etal_2015} requires a starting 3D magnetic cube consistent with a given photospheric magnetogram, which can be obtained by a coronal reconstruction method, for example, from a force-free field extrapolation (potential, linear, or nonlinear). Given that the OVSA spectrum has a complicated shape, contributions from multiple \mw\ sources (a few distinct flaring loops) is expected. In the general case, each flaring loop has its each own twist, which can be captured by a nonlinear force-free field (NLFFF) reconstruction, but not by either potential or linear force-free field (LFFF) extrapolation. NLFFF reconstructions require a vector bottom-boundary condition, which is unavailable at the time of flare, but the line-of-sight magnetograms from SoHO/MDI are available. Thus, only LFFF reconstructions could be performed, which may require using different data cubes (with distinct force-free parameters $\alpha$) to model different co-existing flaring loops, like in the recent study of a cold flare  by \citet{Fl_etal_2016}.

\begin{figure*}\centering
\includegraphics[width=0.96\columnwidth]{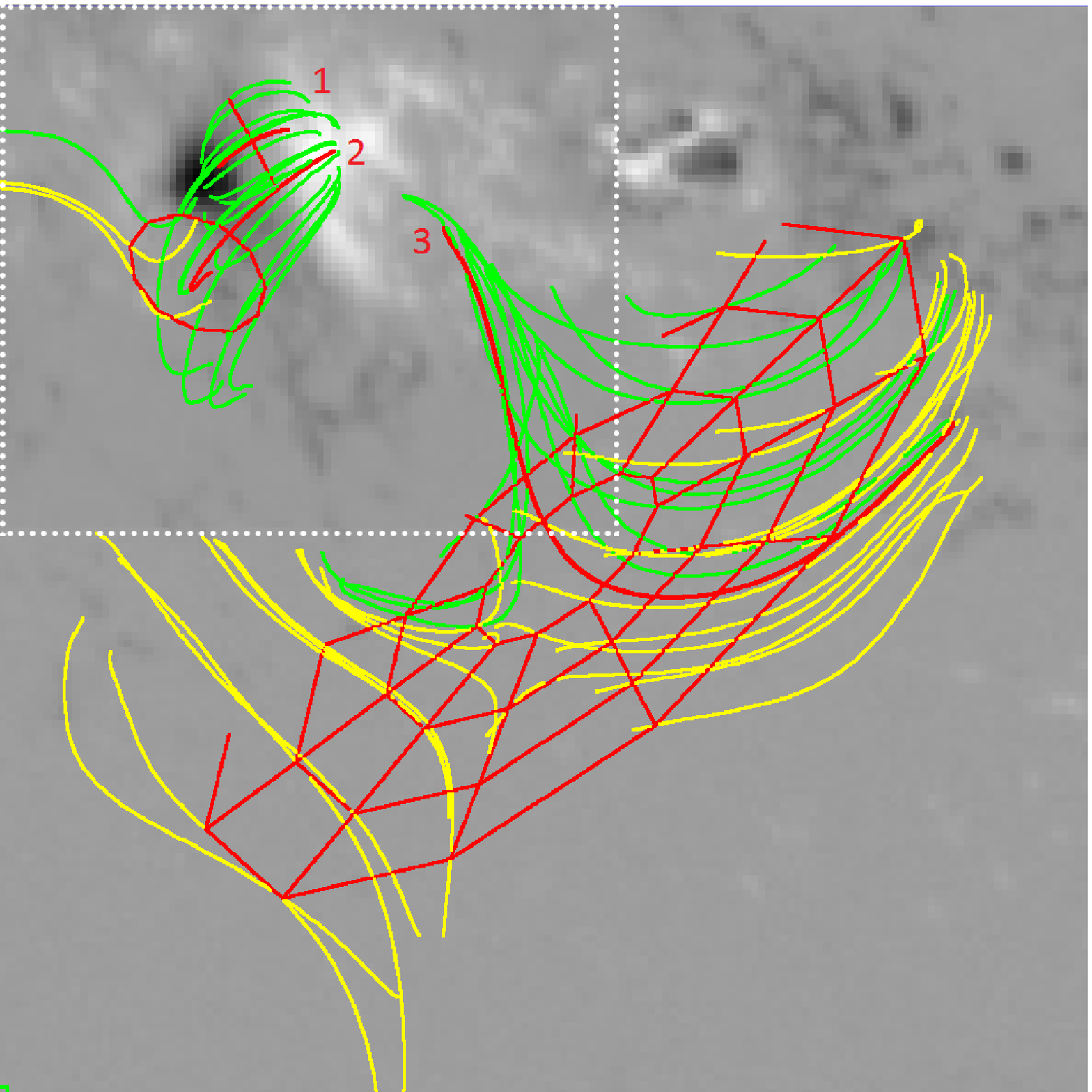}\quad 
\includegraphics[width=0.96\columnwidth]{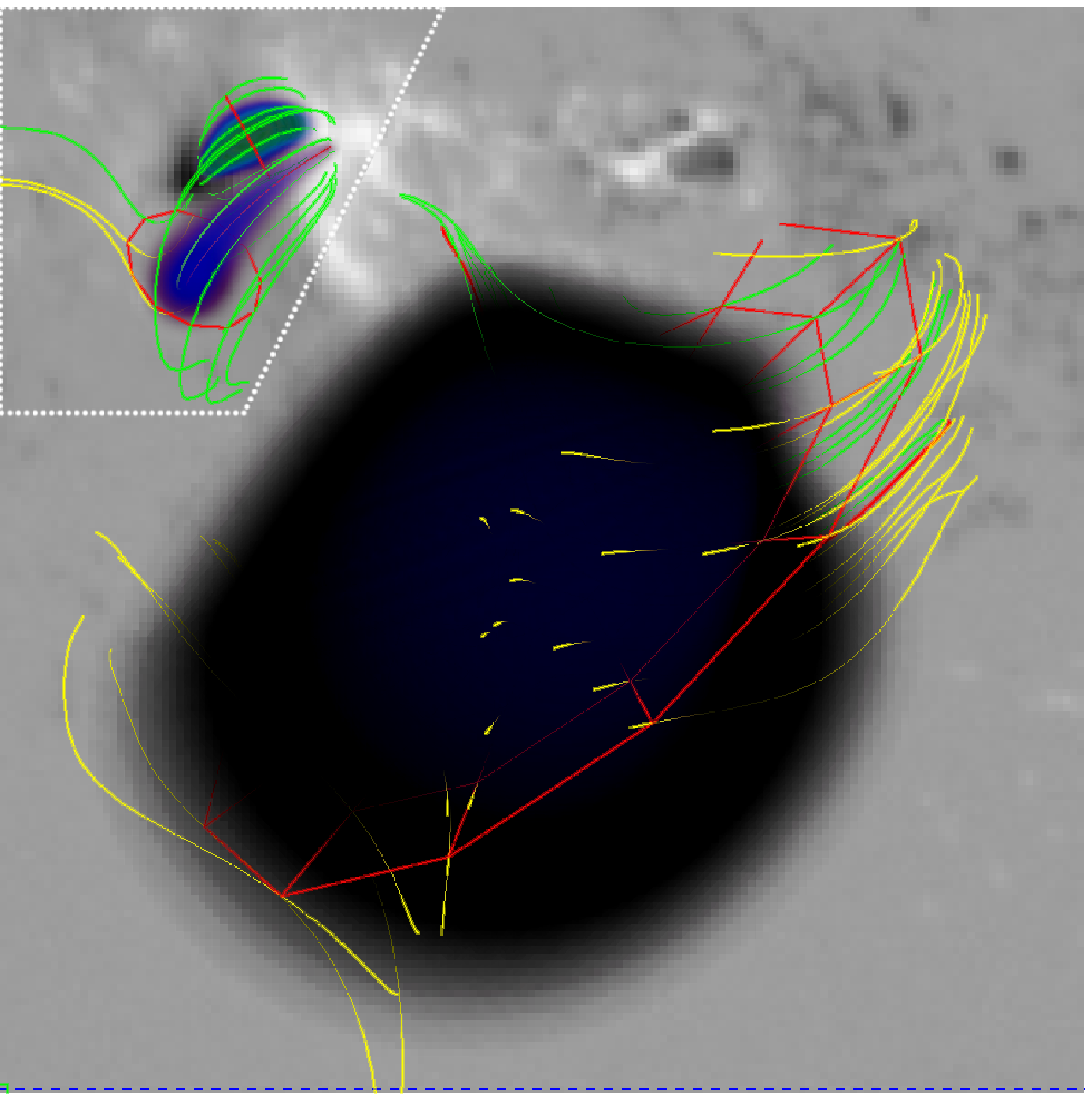}\\
\caption{\label{f_20010825_model}  A composite view of the 3D model developed for the flare. Left: the field lines visualizing three model flux tubes labeled 1, 2, and 3. In each case, the red line shows a corresponding central (reference) field line, while the red circles represent the reference cross-sectional areas of the flux tubes. Closed field lines are shown in green, while the open ones--in read. The bottom image is the SOHO/MDI LOS magnetogram. The full image corresponds to the field of view (FOV) used to develop the `plume' model, while the white dotted lines demarcate the FOV in the upper left corner used to develop the LFFF model for the main (compact) flaring loops. Right: the same model, where the spatial distributions of nonthermal electrons in all the flux tubes are shown by the dark-blue and black volume. The white dotted lines approximately demarcate the areas from which the dominant contribution to the emission comes from either Flux tubes 1--2 or the plume.
}
\end{figure*}

\begin{figure}\centering
\includegraphics[width=0.96\columnwidth]{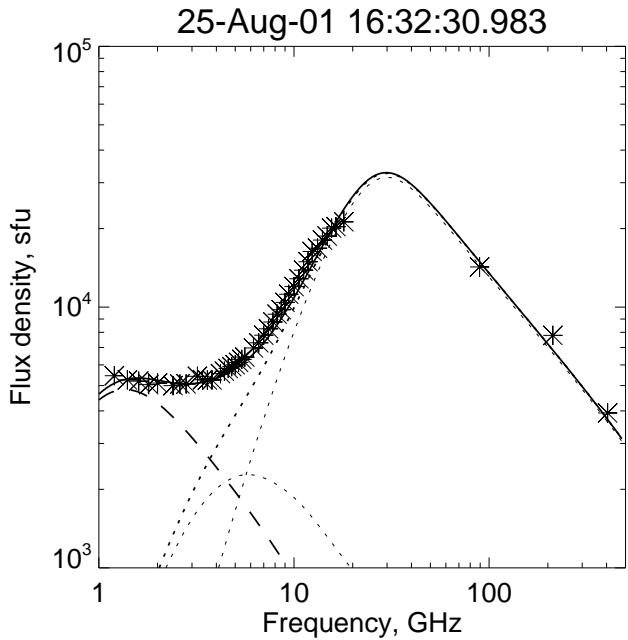}\\
\caption{\label{f_20010825_spec} Synthetic model spectrum (thick solid line) and its components and the observed broad-band radio spectrum (asterisks). The thin dotted lines show contributions from Fluxtubes 1 \& 2, while the thick dotted line--the total spectrum produced by these two fluxtubes. The dashed line shows the spectrum produced by the plume.  
 }
\end{figure}

Here we follow the same modeling strategy as that  described by \citet{Fl_etal_2016} in detail.  We select a single time frame in the early decay phase, $\sim$16:32:31~UT, and start by modeling the high-frequency emission, including the spectral peak. Given that the spectral peak frequency is very high, $f\peak\sim30$~GHz, even in the early decay phase, the \mw\ source must have a strong magnetic field. This agrees with the OVSA imaging data: the high-frequency OVSA images project onto the core of the AR, where the magnetic field is strong. Thus, we attempted a few LFFF extrapolations at an area surrounding the core of the AR and found that the one with $\alpha=2\times10^{-10}$~cm$^{-1}$ offers the connectivity implied by a double Yohkoh 32.7--52.7~keV source, presumably highlighting the footpoints of the main flaring loop. The corresponding model fluxtube (Fluxtube 1) crosses the brightness centers of the high-frequency OVSA images, see Figure~\ref{f_20010825_model} as an example\footnote{Given imperfect nature of the OVSA imaging discussed elsewhere \citep{Fl_etal_2015, Fl_etal_2016_narrow}, we use the OVSA images as general guides only in terms of the source locations and areas, but we do not attempt to reproduce the detailed shapes of the images in our modeling.}. The center field-line of the flux tube is 18.6~Mm long and has a strong looptop magnetic field of 770~G, with stronger footpoint values of $\sim$1350~G and $\sim$2300~G. To roughly match the \mw\ spectral peak and high-frequency slope, we adopted the reference radius of the fluxtube (this reference radius of the circle visualized as a red circle normal to the central field line, also red, is used to normalize the transverse distributions of the thermal plasma and nonthermal electrons) to be roughly 10~Mm at the looptop. We then filled the fluxtube with $N_1\approx1.28\times10^{34}$ nonthermal electrons above 10~keV having a single power-law distribution over energy with the index $\delta_1=2.2$. The effective volume of this fluxtube, defined as $V_1=\left[\int n_b dV\right]^2/\int n_b^2 dV$, where $n_b$ is the number density of nonthermal electrons, is $V_1\approx2.2\times10^{27}$~cm$^{3}$. The \mw\ spectrum produced by this fluxtube is shown by a dotted curve in Figure~\ref{f_20010825_spec}. This spectrum is not sensitive to the thermal plasma density at least if $n_0\lesssim10^{11}$~cm$^{-3}$; thus, the thermal number density in Fluxtube 1 cannot be constrained by the radio data.

The model spectrum is significantly steeper at the low frequencies than the observed one. The reason for such a deviation is well understood: the real flaring volume is considerably more nonuniform than a single flaring loop responsible for the high-frequency \gs\ emission. Therefore, guided by the OVSA images at $f=6-12$~GHz, we have to distribute more nonthermal electrons over the volume surrounding Fluxtube 1. To do so, we create another (asymmetric) fluxtube (\#~2; see Figure~\ref{f_20010825_model}) adjacent to Fluxtube 1, which has a broader range of magnetic field values: the looptop magnetic field is 47~G, while the footpoint values are $\sim$180~G and $\sim$1500~G. The parameters of Fluxtube 2 are not well constrained by the \mw\ data because this flux tube dominates only a very restricted spectral range; so a range of models could be consistent with the data. In one such model the total number of the nonthermal electrons in Fluxtube 2 is comparable to that in Fluxtube~1, $N_2\approx2\times10^{34}$ (the same spectral shape with $E_{\min}=10$~keV and the power-law index $\delta_2=2.2$ are adopted). The effective volume of this fluxtube is $V_2\approx1.32\times10^{28}$~cm$^{3}$. As the thermal number density is varied, we find that for $n_0\gtrsim2\times10^{10}$~cm$^{-3}$ the \mw\ spectrum produced by this source becomes too steep at the low-frequency side to match the observed one, due to so-called Razin suppression \citep{Bastian_etal_1998, victor, Fl_etal_2016_narrow}. However, this is not a stringent constraint given that the parameters of this loop are anyway not well constrained by the data, so any value $n_0\lesssim10^{11}$~cm$^{-3}$ is likely consistent with the data.

\begin{figure*}\centering
	\includegraphics[width=0.98\columnwidth]{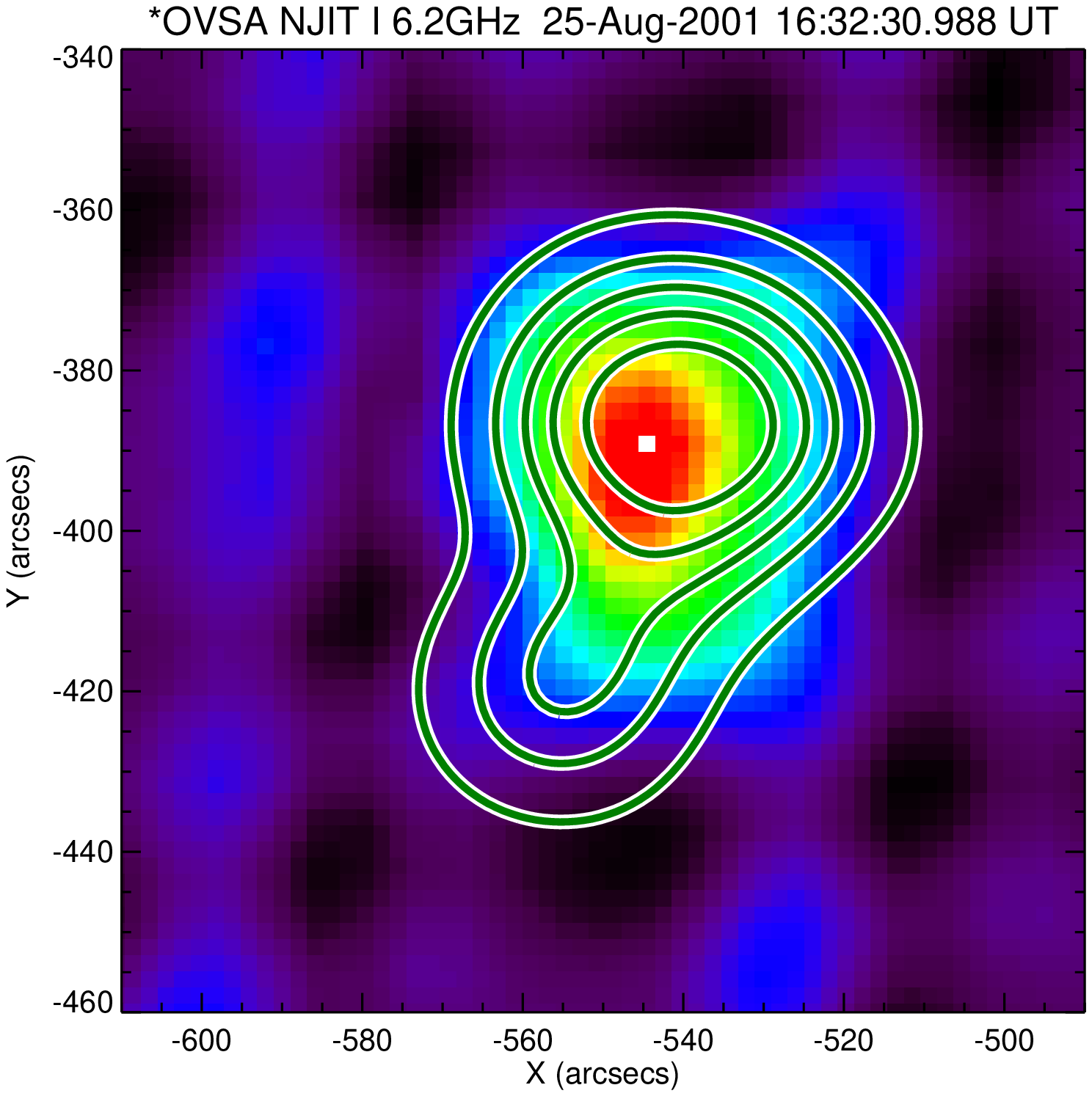}
	\includegraphics[width=0.93\columnwidth]{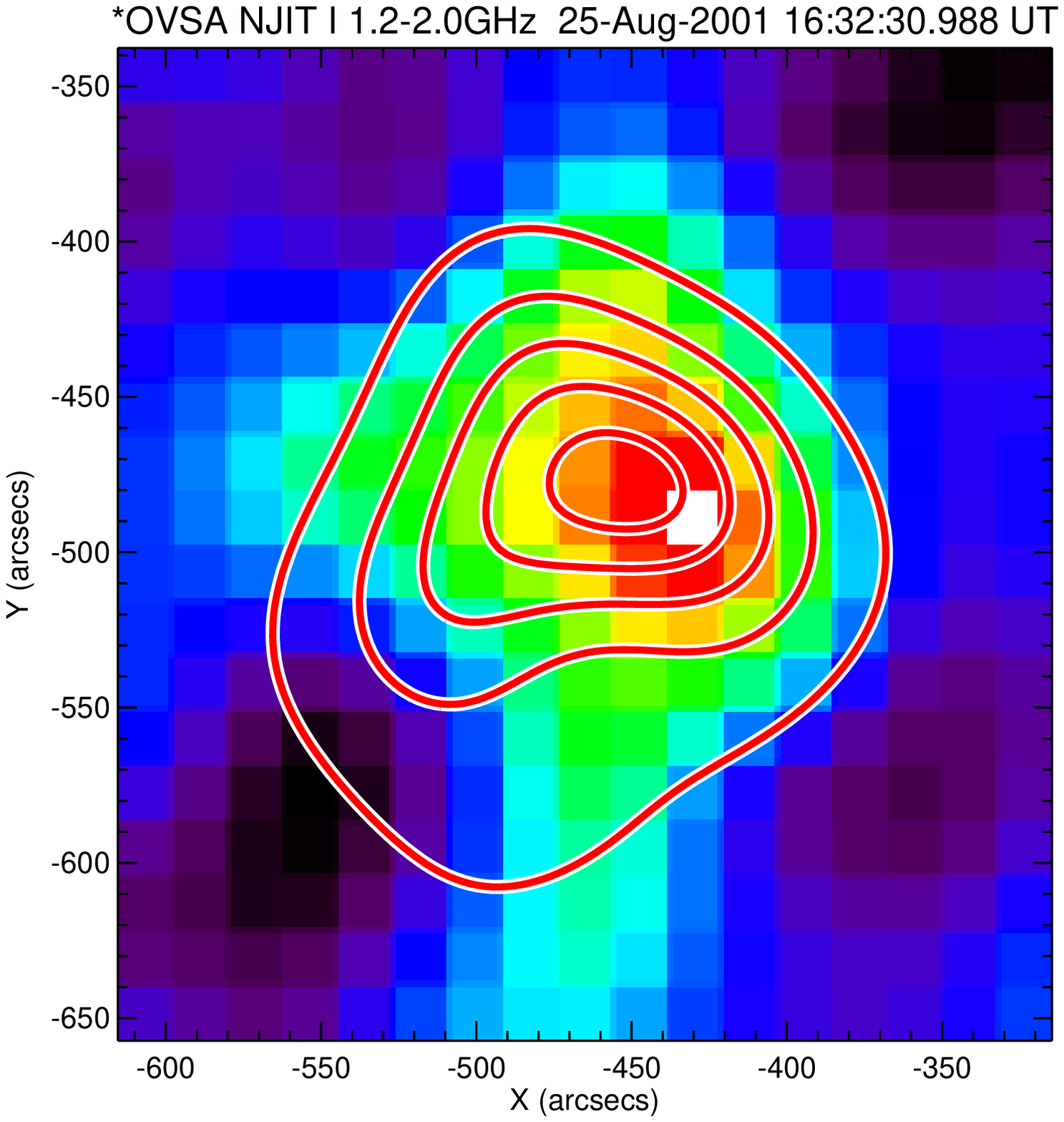}
	\caption{\label{f_20010825_model_map_ON_OVSA_map}
		Left: model \mw\ emission at 6.2~GHz on top of OVSA image at 6.2~GHz. Right: model \mw\ emission at 2~GHz on top of OVSA frequency synthesis image at 1.2--2~GHz. The model images are convolved with the idealized gaussian OVSA beam with the parameters computed for the given date, time, and selected subset of antennas (antenna \#~7 was excluded): $a=  54.87''/f$[GHz], $b=  46.61''/f$[GHz], $\phi= 48.059^\circ$.  $a$ \& $b$ are ellipse semiaxes at FWHMs,  so, the convolution is performed with $\sigma_a=46.6''/f$[GHz] \&  $\sigma_b=39.6''/f$[GHz], where $\sigma_a$  \&  $\sigma_b$ are gaussian variances used for analytical image convolution.
	}
\end{figure*}

Although the total number of nonthermal electrons is comparable in Fluxtubes 1 \& 2, the \mw\ fluxes and spectral shapes, shown by dotted curves in Figure~\ref{f_20010825_spec}, are remarkably different, which is a direct outcome of dissimilar values of the magnetic field in these two fluxtubes. The combination of these two sources improves the spectral match substantially at frequencies above $\sim$6~GHz, and it also agrees well with the OVSA images; see Figure~\ref{f_20010825_model_map_ON_OVSA_map}.

However, in the lower frequency range the observed spectrum is still much flatter than the model spectrum from Fluxtubes 1 \& 2. As we have discussed, the observed flux density at low frequencies requires a very large source of order $3'$. Interestingly, the patches with the opposite magnetic polarities at the AR core are located too close to each other to offer a fluxtube with the required large size (we checked this directly by inspecting the extrapolated data cube). In addition, OVSA images at low frequencies are displaced south-west by roughly $2'$ relative to the main source located at the AR core. Therefore, to create a model of the low-frequency source, we consider a larger field of view centered at the centroid of the low-frequency OVSA images. We expect that the low-frequency radio source must be mainly composed of a  magnetic flux tube (or tubes) reaching to high heights and, thus, a potential extrapolation is expected to work well. {\gf In fact, such extended magnetic structures were proposed long ago \citep{1993SoPh..144..361D} to interpret SEP events with weak impulsive phases of microwave emission.}

The potential extrapolation data cube does offer a magnetic connection between the AR core and a remote patch of the magnetic field. The corresponding magnetic field lines do cross the centroid of the low-frequency source, of which we selected a ``central'' field line that most closely crosses the low-frequency \mw\ source.
The central field line of the flux tube is 198~Mm long and has a weak looptop magnetic field of $\sim$5~G; the footpoint values are $\sim$300~G. The reference radius of this fluxtube is $\sim90$~Mm; it contains both closed and open field. Because of the extremely large size of the fluxtube, its significant displacement compared with the main flare location, and both open and closed magnetic connectivity visualized by a subset of closed (green) and open (yellow) field lines, we call this fluxtube a plume. The model contains a vast number $N\plume\approx2.32\times10^{36}$ nonthermal electrons above 10~keV with a power-law index $\delta\plume=2.4$. The effective volume of this fluxtube is $V\plume\approx8\times10^{29}$~cm$^{3}$, which implies a mean nonthermal electron density $\bar{n}_b=N\plume/V\plume\simeq3\times10^6$~cm$^{-3}$. The \mw\ spectrum produced by this fluxtube is shown by the dashed curve in Figure~\ref{f_20010825_spec}. In contrast to Fluxtubes 1 \& 2, the thermal number density in the plume is stringently constrained by the radio spectrum. Indeed, for a plume magnetic field of about 5~G, the \gs\ emission at $f\sim1$~GHz will be noticeably suppressed by the Razin-effect for a thermal density as low as $n_0\gtrsim10^{8}$~cm$^{-3}$, which we straightforwardly confirmed by recomputing the radio spectrum from the same model, but with an enhanced thermal number density. The absence of the Razin suppression signatures in the observed spectrum implies that the plume density is correspondingly low, $n_0\lesssim10^{8}$~cm$^{-3}$. The total spectrum obtained by adding up the contributions from two extrapolated magnetic cubes used for the modeling is shown by the thick solid line, which is an almost perfect fit to the data. {The only apparent disadvantage of the model visualised in Figure~\ref{f_20010825_model} is the lack of magnetic connectivity between the main flaring loops and the plume. In fact, this is an unavoidable outcome of the oversimplified treatment of the corresponding flaring loops using distinct potential and LFFF magnetic extrapolations, so the compact flux tubes (\#1 and 2) and the plume are isolated from each other. No doubt that the real magnetic field in the flaring volume is more complicated than the modeled one, so the required magnetic connectivity is not unexpected.} Given that the simulated images agree well with the observed ones throughout the entire spectral range for which images are available, see examples in Figure~\ref{f_20010825_model_map_ON_OVSA_map}, we conclude that the developed model is validated by comparison with the data.

\section{Discussion \& Conclusions}

The combination of the \mw\ data and 3D modeling described above offers an apparently unexpected conclusion about the spatial structure of the magnetic flux tubes involved in the 2001-Aug-25 flaring event: in addition to relatively compact structures seen in high-frequency \mw\ and sub-THz emission, as well as in WL, EUV, SXR, and HXR \citep[e.g.,][]{2003ApJ...595..483M, 2004SoPh..223..181R,2006SoSyR..40..104K, 2013PASJ...65S...4G}, there exists a large-scale plume seen only at low radio frequencies. The reason for the plume being undetectable at most of the wavelengths is fully understandable: the number density of the thermal plasma in the plume volume is very low, not higher than $\sim10^8$~cm$^{-3}$; otherwise, the low-frequency radio emission would be strongly suppressed by the Razin-effect, given that the magnetic field is very low in the plume. But by the same token, neither UV nor X-ray emission would be observable, given that they are proportional to the square of the density. The high-frequency radio emission from the plume is also expected to be negligible, given its relatively low magnetic field. At the same time, the optically thick low-frequency \gs\ emission has a large brightness in case of the weak magnetic field; so the very combination of parameters that makes it invisible at high frequencies is favorable for the plume to dominate at low frequencies. This means that having \mw\ imaging observations at low frequencies is critically important, and, perhaps, the only means for detection and quantitative analysis of large-scale structures like the plume detected here, which contains a vast number of  nonthermal particles, and thus is energetically important.

In the current instance, we developed a 3D model based on coronal magnetic field reconstructions obtained using the available photospheric magnetogram. In particular, the full 3D model of the plume was created based on a potential magnetic extrapolation encompassing a relatively large region of interest. One of the outcomes of this model is a determination of the 3D distribution of nonthermal electrons, which could imply that we have full information about the nonthermal electrons present in the plume. However, this particular distribution of nonthermal electrons must be taken with considerable caution  for the following reasons. First, the main contribution to the low-frequency flux density comes from the areas where the magnetic field is rather weak; thus, the energies of the emitting electrons must be proportionally large. Indeed, the brightness temperature of about $3\times10^9$~K implies that electrons with energy $\gtrsim300$~keV are the main contributors to the radio emission; thus, the low-frequency spectrum is almost insensitive to the electrons with lower energies, which makes it difficult to estimate the total number of nonthermal electrons in the plume. And second, given that the low-frequency emission is optically thick, we only observe emission from a layer down to the optical depth around 1, rather than from the entire plume volume, which also increases the uncertainty in the estimate of the nonthermal electron number.

These uncertainties can in principle be removed  when spatially resolved spectra from the plume location are available through true \mw\ imaging spectropolarimetry. The OVSA data were not of sufficient quality to accomplish this, but such capabilities are coming online now from  several new or upgraded \mw\ arrays. Forward fitting of the spatially resolved polarized brightness spectra \citep{Gary_etal_2013} can recover the source parameters more reliably and, thus, will permit more quantitative estimates of the detailed properties of large-scale plumes, including their build-up and evolution.

\acknowledgments

We are thankful to our colleagues V. Kurt for pointing our attention to this event and V. Grechnev for valuable discussions and exchange of the associated data and ideas.
This work was supported in part by NSF grants AGS-1262772  and AST-1615807, and NASA grant NNX16AL67G to New Jersey Institute of Technology.

\bibliographystyle{apj}
\bibliography{WP_bib,solar_radio,Xray_ref,fleishman}

\end{document}